# DEALING WITH MEGAWATT BEAMS*†

N.V. MOKHOV

*Fermilab, MS 220, P.O. Box 500*
*Batavia, IL 60510, U.S.A.*

*Work supported by Fermi Research Alliance, LLC under contract No. DE-AC02-07CH11359 with the U.S. Department of Energy.
†Presented at *10th Workshop on Shielding Aspects of Accelerators, Targets and Irradiation Facilities (SATIF-10),* June 2-4, 2010, CERN, Geneva, Switzerland.

# Dealing with MegaWatt Beams

Nikolai V. Mokhov

Fermi National Accelerator Laboratory, Batavia, IL 60510 USA

## Abstract

*The next generation of accelerators for MegaWatt proton, electron and heavy-ion beams puts unprecedented requirements on the accuracy of particle production predictions, the capability and reliability of the codes used in planning new accelerator facilities and experiments, the design of machine, target and collimation systems, detectors and radiation shielding and minimization of their impact on environment. Recent advances in code developments are described for the critical modules related to these challenges. Examples are given for the most demanding areas: targets, collimators, beam absorbers, radiation shielding, induced radioactivity and radiation damage.*

## Introduction

The next generation of accelerators for MegaWatt proton, electron and heavy-ion beams moves us into a completely new domain of extreme specific energies of ~0.1 MJ/g and specific power up to 1 TW/g in beam interactions with matter. Challenges arise also from increasing complexity of accelerators and experimental setups, as well as design, engineering and performance constraints. All these put unprecedented requirements on the accuracy of particle production predictions, the capability and reliability of the codes used in planning new accelerator facilities and experiments, the design of machine, target and collimation systems, detectors, radiation shielding and minimization of their impact on environment. This leads to research activities involving new materials and technologies, and also code developments whose predictive power and reliability being absolutely crucial.

A list of high-power proton and heavy-ion accelerators includes those under

- Operation: ISIS, PSI, J-PARC and SNS with 0.2 to 1 MW beam power and upgrade plans up to 1-3 MW.
- Construction: CSNS, 0.1-0.2 MW.
- Design: FAIR and FRIB with proton to uranium beams up to 0.4 MW, ESS with a few MW beams, and Project-X with proton beams up to 4 MW.
- Consideration: subcritical accelerator-driven systems with proton beams up to 10 MW.

Another category is high-energy colliders, operating: proton/antiproton 2 MJ beams of the Tevatron, and up to 350 MJ proton beams of the LHC, and planned: $e^+e^-$ (ILC and CLIC, up to 20 MW) and $\mu^+\mu^-$ with a 4-MW proton source.

1. **Critical Areas**

Components of the three critical systems of the accelerators listed are especially vulnerable to the impact of the high-power beams: target stations, beam absorbers and collimators.

*Targets*

The principal issues include: production and collection of maximum numbers of particles of interest; suppression of background particles in the beamline; target and beam window operational survival and lifetime (compatibility, fatigue, stress limits, erosion, remote handling and radiation damage); protection of focusing systems including provision for superconducting coil quench stability; heat loads, radiation damage and activation of components; thick shielding and spent beam handling; prompt radiation and ground-water activation. The most challenging one is a choice of a target technology. Fundamentally, it depends on a peak power density and power dissipation in the target material as illustrated in Fig. 1 generated for the FRIB project [1]. It is clear how severe this problem is for intense heavy-ion beams (up to uranium) of a small spot size impinging on a target or beam window material. As an example of this extreme, the uranium ion beam power at SIS-100 of the FAIR project [2] will be up to 0.1 TW, ion energy range 0.4 to 27 GeV/u, peak specific energy and power in a lead target 0.1 MJ/g and 1 TW/g, respectively. Rotating multi-slice carbon disks, liquid lithium or lead targets for heavy ions and open mercury jets for proton beams at neutrino factories, no-material windows are just several of technologies considered in such projects. Radiation damage to solid targets, downstream magnets and auxiliary equipment is identified as one of the key issues in these systems along with reliability and cost of complex remote handling equipment.

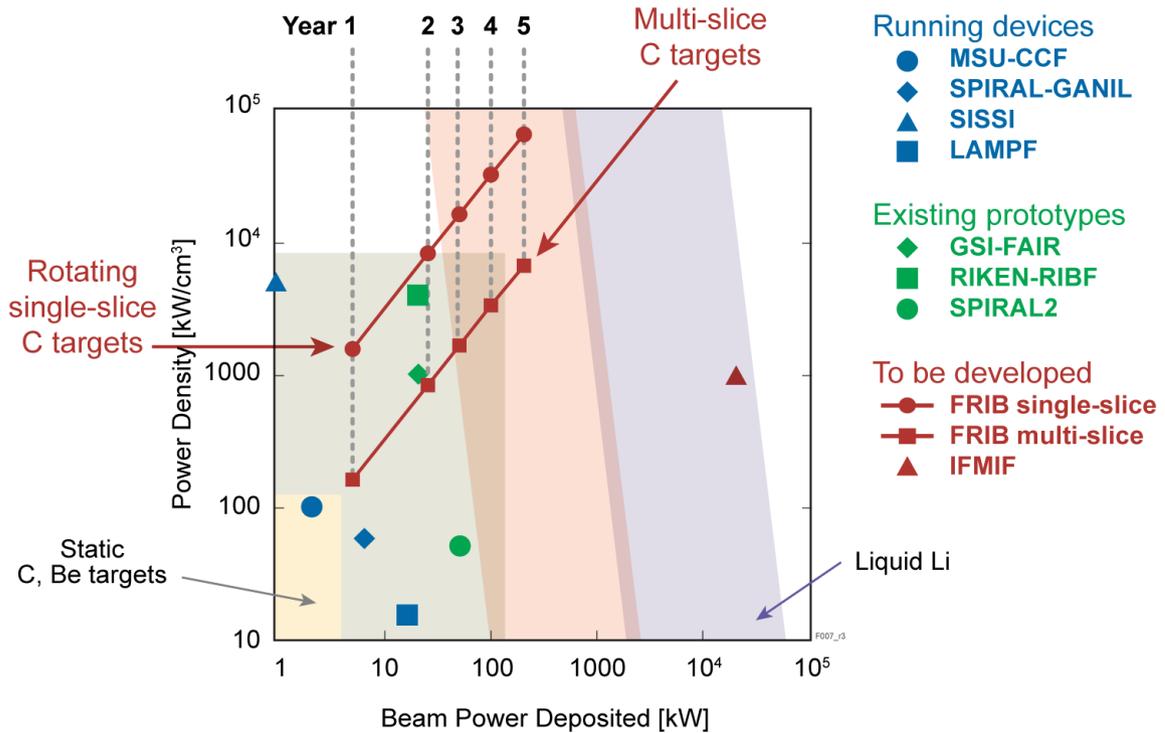

**Figure 1:** Choice of target technology (Courtesy: W. Mittig)

*Beam Absorbers*

Absorbers for misbehaved beams along the beamlines, abort beam dumps and those downstream of the production targets and interaction regions (at linear colliders) are another challenging systems in the MegaWatt accelerators. These should be able to withstand an impact of beams of up to full power, say, 0.2 to 20 MW, without destruction over a designed life-time (at least a few years), fully contain the beam energy, and execute the initial shielding functions. The absorber technologies for high-intensity beams include:

- A laminated graphite core in a cooled aluminum shell. It is proven in more than 20-years of an operational experience at the Tevatron, with the peak instantaneous temperature rise of $\Delta T$ ~1000 °C per pulse. The core is contained in steel shielding surrounded by concrete. A similar design is used at the LHC with the beam swept in a spiral during the abort pulse.

- A stationary beryllium, aluminum or nickel wall liquid-cooled dump. Thin two-layer walls are arranged in a V-shape, with a liquid flowing between the layers and a beam hitting the walls at a small grazing angle. It is successfully used in several high-power proton machines, but is not feasible for intense heavy-ion beams: estimated life-time due to radiation damage is only 3 months for 0.4 MW uranium beam at FRIB [1].

- A water-cooled aluminum-shell rotating drum considered for the FRIB project, with a life-time estimated as 5 years.
- A water-vortex beam absorber considered for an 18-MW electron beam at ILC. The beam is rastered with dipole coils to avoid water boiling. The entrance beam window and catalytic recombination are of a serious concern in this design.

*Collimators*

Only with a very efficient beam collimation system can one reduce uncontrolled beam losses in the machine to an allowable level, thus protect personnel and components, maintain operational reliability over the life of the machine, provide acceptable hands-on maintenance conditions, and reduce the impact of radiation on environment, both at normal operation and accidental conditions. Collimators – as the last line of defense in high-power accelerators - must withstand a predefined fraction of the beam hitting their jaws and - at normal operation - survive for a time long enough to avoid very costly replacements. Design of the collimation systems is especially challenging at the high-energy colliders. At the LHC, the overall collimation efficiency should be better than 99.9% (this value is typical of the Tevatron collider). The system should manage substantial beam losses of 0.5 MW at normal (slow) operation, 20 MW in a transient regime of ~1 ms, and up to 5 TW at a beam accident (1 MJ in 0.2 $\mu$s into 0.2 mm$^2$ area). Novel collimation techniques include crystal channeling and multiple volume reflection, hollow electron beam scraper, volume reflection radiation, rotatable/consumable collimators, and marble shells to mitigate hands-on maintenance problems.

## 2. Modeling Challenges

Particle transport simulation tools and the physics models and calculations required in developing relevant codes are all driven by these demanding applications. This puts unprecedented requirements on the accuracy of particle production predictions, the capability and reliability of the codes used. The challenge is detailed and accurate (to a % level) modeling of all particle interactions with 3-D system components (up to tens of kilometers of the lattice in some cases) in energy region spanning up to 15 decades as a basis of accelerator, detector and shielding designs and their performance evaluation, for both short-term and long-term effects.

Five general-purpose, all-particle codes are capable of this - FLUKA, GEANT, MARS, MCNPX and PHITS - and they are used extensively worldwide for accelerator applications. A substantial amount of effort (up to several hundreds of man-years) has been put into development of these codes over the last few decades. The user communities for the codes reach several thousands of people worldwide. All five codes can handle very complex geometry, have powerful user-friendly built-in GUI interfaces with magnetic field and tally viewers, and variance reduction capabilities. Tallies include volume and surface distributions (1D to 3D) of particle flux, energy, reaction rate, energy deposition, residual nuclide inventory, prompt and residual dose equivalent, displacement-per-atom (DPA) for radiation damage, event logs, intermediate source terms, etc.

As an example, the advanced features in the MARS15 code [3] – instigated by accelerator developmental needs – include: reliable description of cross-sections and particle yields from a fraction of eV to many TeV for hadron, photon and heavy-ion projectiles (event generators); precise modeling of leading particle production and low-momentum transfer processes (elastic, diffractive and inelastic), crucial for beam-loss and collimation studies; reliable modeling of $\pi^0$-production (electro-magnetic showers), $K^0$-production (neutrino and kaon rare decay experiments), proton-antiproton annihilation, and stopped hadrons and muons; nuclide inventory, residual dose, DPA, hydrogen and helium production; precise modeling of multiple Coulomb scattering with projectile and target form-factors included; reliable and CPU-efficient modeling of hadron, lepton and heavy-ion electromagnetic processes with knock-on electron treatment and - at high energies – bremsstrahlung and direct pair production; hadron/muon photo- and electro-production; accurate particle transport in arbitrary geometry in presence of magnetic fields with objects ranging in size from microns to kilometers; variance reduction techniques; enhanced tagging of origin of a given signal/tally – geometry, process and phase-space – invaluable for source term and sensitivity analyses; user-friendly geometry description and visual editing; interfaces to MAD, ANSYS and hydrodynamics codes.

An example of a state-of-the-art particle production module in MARS15 [3] is the Quark-Gluon String Model event generator LAQGSM [4] combined with the Fermi break-up model, the coalescence model, and the generalized evaporation-fission model. The module is used for photon, hadron and heavy-ion projectiles with energies ~1 MeV/A to 1 TeV/A. This provides a powerful fully theoretically-consistent modeling of exclusive and inclusive distributions of secondary particles, spallation, fission, and fragmentation products needed for numerous applications.

### 3. Opportunities and Benchmarking for High-Power Beam Applications

All new developments in the codes are always a subject for thorough benchmarking [5], which is crucial for applications in the MegaWatt domain. Two areas relevant to the SATIF mission are of a special importance here: beam impact on materials and thick-shielding applications.

As stated above, the short- and long-term effects in materials impacted by high-intensity beams have a potential of a showstopper in many instances, with a choice of target, beam window, collimator and beam absorber technologies being the most challenging. Corresponding R&D is launched in all the projects listed in Introduction. The analysis performed at the dedicated Materials Workshop [6] revealed that despite the difference in energy and other beam parameters, all the high-intensity beam facilities have the common issues in this area: choice of the most promising materials, the most suitable configurations of target, collimator and absorber assemblies, materials and technological limits, the critical properties of materials for a specific system, and – in many cases – a lack of reliable experimental data matching the needs of the new generation of accelerators.

The needs for measurements on materials include: mechanical properties for different doses and different strain rates; electrical and thermal properties for different radiation doses; damage thresholds and extent; radiation resistance; desorption and vacuum properties versus temperature. The following measurements are needed for the mockup assemblies: robustness against beam shock impact; cooling efficiency; pressure increase in cooling water pipes; geometrical stability (flatness, deformations, etc.); impedance, RF trapped modes and vibrations. The facilities to perform these tests are available around the world; the issue is to match their capabilities to the conditions required.

An analysis of displacements per target atom (DPA) is the most universal way to describe deterioration of material critical properties under irradiation. Predictions of the MARS15 DPA model have been recently compared to other calculations [7]. The first case is a 1-GeV proton beam of 1-cm$^2$ area on a 3-mm thick iron target. SRIM, PHITS and MCNPX results are courtesy of Susana Reyes. As one can see in Table 1, there is a quite substantial difference between the predictions, with SRIM giving a very small value and the MARS15 result being a factor of 2.6 to 2.9 above those by PHITS and MCNPX. Calculated with MARS15 contributions to DPA of physics processes are as follows: 75.5% nuclear inelastic, 16% nuclear elastic, 2.75% electromagnetic elastic, 5.5% low-energy neutrons, and 0.25% electrons. The dominance of nuclear interactions in this case explains the above differences.

Table 1. DPA for 1-GeV protons on 3-mm iron.

| Code | SRIM | PHITS | MCNPX | MARS15 |
|---|---|---|---|---|
| DPA/pot | 1.18e-22 | 2.96e-21 | 3.35e-21 | 8.73e-21 |

The second case is a 0.32-GeV/u Uranium beam of 9-cm$^2$ area on a 1-mm thick beryllium target. The SRIM and PHITS results are again a courtesy of Susana Reyes. Table 2 shows that the SRIM and MARS15 results are now very close to each other, while those calculated with PHITS are a factor of 70 lower. Calculated with MARS15 contributions to DPA of physics processes are as follows: 0.3% nuclear inelastic, 99.06% electromagnetic elastic, 0.02% low-energy neutrons, and 0.62% electrons. The dominant role of Coulomb scattering in this case explains the similarity of the SRIM and MARS15 predictions.

Table 2. DPA for 0.32-GeV/u Uranium on 1-mm beryllium target.

| Code | SRIM | PHITS | MARS15 |
|---|---|---|---|
| DPA/pot | 2.97e-20 | 5.02e-22 | 2.13e-20 |

The third case is a 0.13-GeV/u Germanium beam of 0.004-cm$^2$ area on a 1.2-mm thick tungsten target. The TRIM and PHITS results are a courtesy of Yosuke Iwamoto. Table 3 gives calculated DPA values in the first hundred microns of the target. The difference between TRIM and MARS15 needs to be understood.

Table 3. Entrance DPA for 0.13-GeV/u Germanium on 1.2-mm tungsten target.

| Code | TRIM | PHITS | MARS15 |
|---|---|---|---|
| DPA/pot | 8.04e-16 | 1.25e-17 | 1.43e-16 |

A majority of data on radiation damage is available for reactor neutrons. Studies with hundred MeV protons [8] have revealed that a threshold of about 0.2 DPA exists for carbon composites and graphite. MARS15 studies helped realize that the BLIP beam tests with 0.16-0.18 GeV protons can emulate the 2.3-MW LBNE neutrino target situation for a 120-GeV proton beam (Table 4). It turns out that despite a substantial difference in the beam energies in these cases, nuclear interactions and Coulomb scattering contribute about the same way (45-50% each) to the peak DPA in thick graphite targets irradiated at these two facilities. A corresponding experiment at BLIP with a 0.18-GeV proton beam of 96-mA current on a set of carbon-based materials has been launched earlier this year with post-irradiation tests started in late summer 2010.

Table 4. Peak DPA in POCO graphite targets at BLIP and LBNE.

| Target | $E_p$ (GeV) | Beam $\sigma$ (mm) | $N_p$ (1/yr) | DPA (1/yr) |
|---|---|---|---|---|
| LBNE | 120 | 1.1 | 4.0e20 | 0.45 |
| BLIP | 0.165 | 4.23 | 1.12e22 | 1.5 |

A Japan-USA collaboration (JASMIN) conducts – since 2007 – joint studies at Fermilab facilities focusing on issues related to high-intensity accelerators [9]: measurements and code benchmarking in the deep-penetration shielding conditions; studying effects of shielding composition and imperfections; thick target particle yields; nuclide production; residual activation; characterization of radiation fields for studies of radiation effects in materials and electronics components. One of the unique components of this all-encompassing program is measurements of material activation due directly to muon-induced photo-spallation and due to hadrons produced by muon interactions in shielding that can become quite important at high beam power hadron and lepton colliders and neutrino factories.

4. Summary

At the new generation of accelerators, extremely high peak specific energy (up to ~0.1 MJ/g) and specific power (up to ~1 TW/g) in beam interactions with matter make design of such critical systems as targets, absorbers and collimators very challenging, requiring a novel approach. This also puts unprecedented requirements on the accuracy, capability and reliability of the simulation codes used in the designs. Particle production, DPA, nuclide inventory, deep-penetration, energy deposition and hydrodynamics coupling are the modules of special importance. Benchmarking in these areas is absolutely crucial. Justified emulation of the extreme conditions at the existing lower energy and beam power facilities is the way to go. The JASMIN (Fermilab/Japan), BLIP (BNL) and HiRadMat (CERN) activities are the excellent examples. Joint efforts with material experts are needed.